\documentclass[prd,aps,a4paper,twocolumn,superscriptaddress,nofootinbib,notitlepage]{revtex4-1}

\usepackage{amsmath,amssymb,amsthm,wasysym,amsfonts}
\usepackage{graphicx}
\usepackage{psfrag}
\usepackage{epstopdf}
\usepackage{color}
\usepackage{mathrsfs}
\usepackage{fixmath}

\DeclareSymbolFontAlphabet{\mathrsfs}{rsfs}
\DeclareMathAlphabet\mathbfcal{OMS}{cmsy}{b}{n}

\newcommand{\be}{\begin{equation}}
\newcommand{\ee}{\end{equation}}
\newcommand{\bea}{\begin{eqnarray}}
\newcommand{\eea}{\end{eqnarray}}
\newcommand{\bel}{\begin{align}}
\newcommand{\eel}{\end{align}}

\def\GMc2{G M_{\odot} c^{-2}}

\def\lm{{\ell m}}
\def\tmg{ t^{\rm mrg} }
\def\teobLR{{t^{\rm EOB}_{\Omega \, \rm peak}}}

\def\tnrLR{{t_{\rm extr}^{\rm NR}}}
\def\lm{{\ell m}}
\def\lmax{\ell_{\rm max}}

\def\v{v_\varphi}

\def\lm{{\ell m}}

\def\ii{{\rm i}}
\def\l{{\ell }}

\def\a{{\hat{a}}}

\DeclareSymbolFontAlphabet{\mathrsfs}{rsfs}
\DeclareMathAlphabet{\mathcal}{OMS}{cmsy}{m}{n}

\usepackage{color}

\definecolor{cyan}{rgb}{0,0.9,0.9}
\definecolor{orange}{rgb}{0.9,0.5,0}
\definecolor{magenta}{rgb}{1,0,1}
\definecolor{purple}{rgb}{0.8,0.4,0.8}
\definecolor{gray}{rgb}{0.8242,0.8242,0.8242}

\newcommand{\ihes}{Institut des Hautes \'Etudes Scientifiques, 91440 Bures-sur-Yvette, France}
\newcommand{\lmpt}{Laboratoire de Math\'ematiques et de Physique Th\'eorique,\\ Univ. F. Rabelais - CNRS (UMR 7350), F\'ed. Denis Poisson, 37200 Tours, France}

\begin{document}

\title{Merger states and final states of black hole coalescences: \\ a numerical-relativity-assisted effective-one-body approach}

\author{Thibault \surname{Damour}}
\author{Alessandro \surname{Nagar}}
\affiliation{\ihes}
\author{Lo\"\i c \surname{Villain}}
\affiliation{\lmpt}

\begin{abstract}
We study to what extent the effective-one-body description of the dynamical state of a nonspinning, 
coalescing binary black hole (considered either at merger, or after ringdown) agrees with numerical relativity results.
This comparison uses estimates of the integrated losses
of energy and angular momentum during ringdown, inferred from recent numerical-relativity data. 
We find that the values, predicted by the effective-one-body formalism,
 of the energy and angular momentum of the system agree at the per mil level with their numerical-relativity counterparts, 
 both at merger and in the final state.  This gives a new confirmation of the ability of effective-one-body theory to 
 accurately describe the dynamics of binary black holes even in the strong-gravitational-field regime.
  Our work also provides predictions (and analytical fits)
 for the final mass and the final spin of coalescing black holes for all mass ratios.
\end{abstract}

\date{\today}


\maketitle

\section{Introduction}

The effective one body (EOB) formalism \cite{Buonanno:1998gg,Buonanno:2000ef,Damour:2000we,Damour:2001tu,Buonanno:2005xu,Damour:2008gu}  has been introduced as a new analytical method for describing both 
the dynamics and the gravitational
radiation of coalescing black-hole (BH) binaries during the entire coalescence process, from early inspiral to
ringdown, passing through late inspiral, plunge and merger (see \cite{Damour:2012mv} for a recent EOB review). 
The EOB method started its development in 2000 (before the availability of numerical relativity (NR) simulations of 
coalescing BH's), and made several quantitative and qualitative predictions concerning the coalescence dynamics.
These predictions have been broadly confirmed by  subsequent NR simulations of merging BH's that started
yielding reliable results after the NR breakthroughs of 2005 \cite{Pretorius:2005gq,Campanelli:2005dd,Baker:2005vv}.
[See \cite{Pretorius:2007nq,Pfeiffer:2012pc} for reviews of NR results, and \cite{Mroue:2013xna} for a recent 
impressive example of NR achievements]. As soon as NR simulations became available, a fruitful {\it synergy} between
EOB and NR developed \cite{Buonanno:2006ui,Pan:2007nw,Buonanno:2007pf,Damour:2007vq,Damour:2008te}. This synergy allowed one to complete the EOB formalism by incorporating some crucial
{\it nonperturbative information} contained in NR results. This led to  the definition of improved, NR-informed  versions of 
the EOB formalism,  sometimes referred to as EOBNR or EOB[NR] (see \cite{Damour:2009kr,Buonanno:2009qa,
Pan:2009wj,Pan:2011gk,Taracchini:2012ig,Damour:2012ky}).

Besides providing accurate gravitational waveforms (of direct interest for current gravitational wave detectors), 
the EOB formalism also gives a description of the {\it dynamics} of coalescing binaries. Several aspects of the EOB
dynamical description have been (successfully) compared to NR results, notably:  (i)  gravitational recoil during 
coalescence ~\cite{Damour:2006tr}, (ii)  the spin parameter of the final BH ~\cite{Damour:2007cb}, (iii)   periastron 
advance during inspiral ~\cite{LeTiec:2011bk}, and (iv) the functional relation between energy and angular momentum 
during inspiral~\cite{Damour:2011fu}. The aims of the present work are, on the one hand, to update  the early 
EOB-based study  of the final spin of a (nonspinning) coalescing binary~\cite{Damour:2007cb}, and, on the other hand, 
to complete it by considering the dynamical characteristics (energy and angular momentum) both of the {\it merger state} 
and of the {\it final state} of the BH system. [This aspect of our work is similar to the EOB/NR comparison done in \cite{Damour:2011fu}.]
On the EOB side
our  work will use the latest version of the (nonspinning, NR-informed) EOB formalism~\cite{Damour:2012ky,eobihes}, 
while, on the NR side, it will use the results of  the most recent NR simulations of merging, unequal-mass BH 
binaries~\cite{Scheel:2008rj,Buchman:2012dw,Lousto:2010qx,Lousto:2010ut,Lousto:2010xk,Bernuzzi:2011aj}. 

The paper is organized as follows: In Sec. II we compute from NR results the ringdown losses of energy and 
angular momentum.  In Sec.~III we compare the EOB prediction for the dynamical state of the BH system at 
merger to the corresponding state inferred from NR results, while Sec.~IV  performs such an EOB/NR comparison for the final
dynamical state. Some conclusions are presented in Sec.~V. We use geometric units $G=c=1$. The letter $q$ is used
to denote the mass ratio $q:=m_2/m_1 \geq 1$, while $\nu$ denotes the symmetric mass ratio 
$\nu := m_1 m_2/(m_1+m_2)^2 = q/(q+1)^2$.

\section{NR-computed ringdown losses of energy and angular momentum}

Throughout this paper we shall (conventionally) call ``merger" the moment $\tmg$  when the modulus of the 
$\lm = 2 2$ strain waveform $h_{22}(t)$ [or, equivalently, its Zerilli-normalized version $\Psi_{22}(t)= (R/M) h_{22}/\sqrt{24}$] 
reaches its maximum. We correspondingly refer to the phase following the merger, i.e. $t  >  \tmg$, as being 
the ``ringdown". We shall apply these definitions both to the NR waveform, and to the EOB one. In other words, the 
NR merger moment $\tmg_{\rm NR}$ is computed by considering the maximum of the modulus of  the NR waveform 
as a function of the NR time scale, while the EOB merger moment $\tmg_{\rm EOB}$ is computed from the maximum 
of the EOB waveform as a function of the EOB time scale. In the present Section we only consider the NR waveform, 
and shall estimate the (NR) losses (in the form of fluxes at infinity) of energy and angular momentum during the NR 
ringdown, $ t_{\rm NR} > \tmg_{\rm NR}$. [For brevity, we suppress in the formulas of this Section the label NR on
all the quantities.]

 Using the Zerilli-normalization of the waveform 
$\Psi_{\lm}(t)= (R/M) h_{\lm}/\sqrt{(\l+2)(\l+1) \l (\l-1)}$ the fluxes of energy and angular momentum at infinity are given by

\begin{align}
\label{eq:Edot}
\dot{E}_{(\ell_{\rm max})} &=\dfrac{1}{16\pi}\sum_{\ell =2}^{\ell_{\rm max}}\sum_{m=-\l}^{\ell}
\dfrac{(\ell+2)!}{(\ell-2)!}|\dot{\Psi}_\lm|^2 , \\ 
\label{eq:Jdot}
\dot{J}_{(\ell_{\rm max})} &=-\dfrac{1}{8\pi}\sum_{\ell =2}^{\ell_{\rm max}}\sum_{m=1}^{\ell} m
\dfrac{(\ell+2)!}{(\ell-2)!}\Im\left[\dot{\Psi}_\lm
  \Psi^{*}_\lm \right] \ ,
\end{align}
where the summation over $\l$ extends only up to some maximum value $\ell_{\rm max}$ and 
$\Psi^{*}_{\lm}=(-1)^{m}\Psi_{\ell,-m}$. The corresponding (NR) ringdown losses are then obtained 
by integrating the fluxes~\eqref{eq:Edot},~\eqref{eq:Jdot} over  $ t> \tmg$, namely

\begin{align}
\label{eq:DeltaE}
 E^{\rm rgd}_{(\ell_{\rm max})} & = \int\limits_\textrm{merger}^\infty dt \,\dot{E}_{(\ell_{\rm max})}  \,= \sum_{\ell =2}^{\ell_{\rm max}}  E^{\rm rgd}_\l\, ,\\ 
\label{eq:DeltaJ}
J^{\rm rgd}_{(\ell_{\rm max})} & = \int\limits_\textrm{merger}^\infty dt \,\dot{J}_{(\ell_{\rm max})} \,= \sum_{\ell =2}^{\ell_{\rm max}}  J^{\rm rgd}_\l\, \, .
\end{align}
Here the last expressions on the right denote the decomposition of the integrated ringdown losses into individual
multipolar contributions, obtained by fixing $\l$ and summing over $m$ in Eqs.~\eqref{eq:Edot}, \eqref{eq:Jdot}.
[Note the notational difference between, say, the individual contribution $ E^{\rm rgd}_\l$ and the result of its summation
 up to $\lmax$: $E^{\rm rgd}_{(\ell_{\rm max})}$.]

\begin{figure*}[t]
 \includegraphics[width=0.45\textwidth]{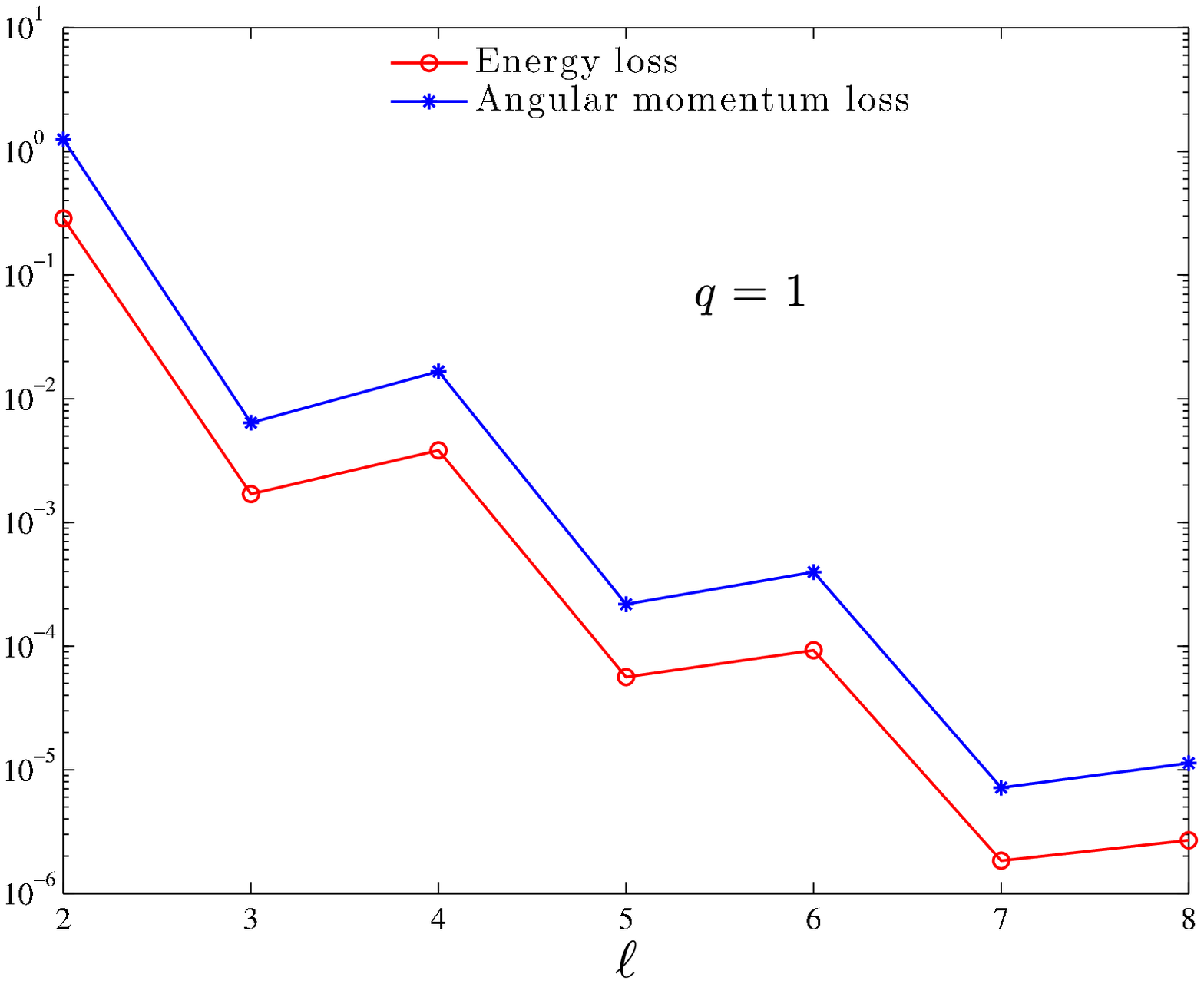}
 \hspace{5mm}
 \includegraphics[width=0.45\textwidth]{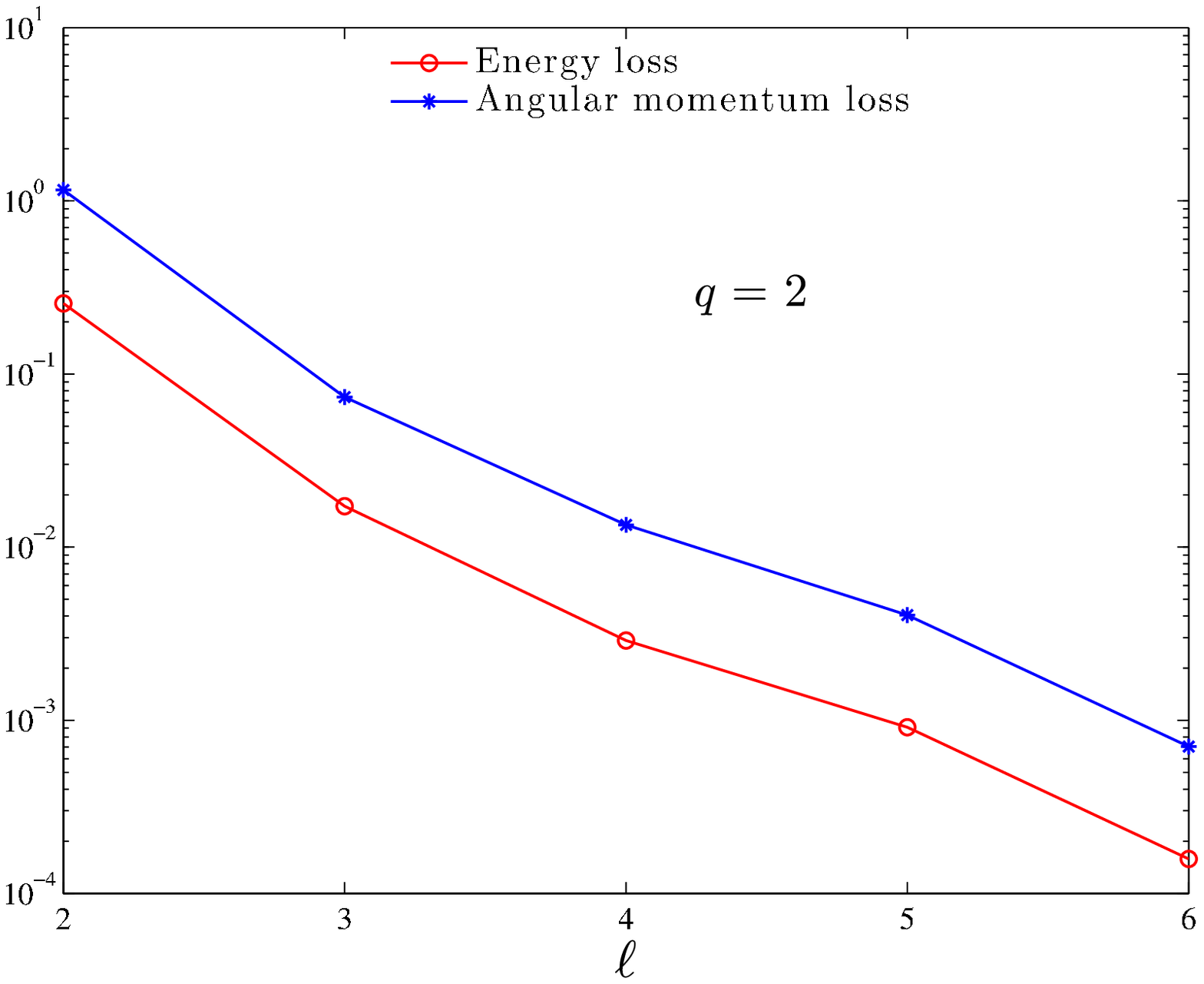}
    \caption{\label{fig:q12} Logarithms of the $\l$-th multipolar contributions, $ E^{\rm rgd}_\ell$ , $J^{\rm rgd}_\ell$ ,
    to the ringdown losses, as functions of $\ell$ for $q=1$ (left) and $q=2$ (right),  exhibiting different behaviors 
    for even and odd values of $\l$; the difference becomes  less pronounced as $q$ increases.} 
\end{figure*}

The NR waveform data available to us have been computed with the SpEC code by the 
Caltech-Cornell-CITA collaboration~\cite{Scheel:2008rj,Buchman:2012dw}. For the
mass ratio  $q=m_2/m_1=1$~\cite{Scheel:2008rj}, all waveform multipoles up to $\lmax=8$ are present. 
When $q\geq 2$~\cite{Buchman:2012dw}, only the dominant $\lm$ multipoles up to $\lmax=6$ are available, 
namely $\lm=(22,21,33,31,44,55,66)$. Starting from this information we tried 
to improve the determination of the ringdown losses  
$ E^{\rm rgd}_{(\ell_{\rm max})} ,  J^{\rm rgd}_{(\ell_{\rm max})} $ by studying the decay of the partial multipolar contributions   
$ E^{\rm rgd}_\l ,  J^{\rm rgd}_\l $  entering the above sums as $\l$ increases.  When $q \leq 2$
we found that the decay, as  $\l$ increases, of  $ E^{\rm rgd}_\l ,  J^{\rm rgd}_\l $ is {\it non monotonic}. 
More precisely, we found that, when  $q \leq 2$,  the partial multipolar contributions $ E^{\rm rgd}_\l ,  J^{\rm rgd}_\l $ 
corresponding to even or odd values  of $\l$  behave in different ways. However, if we compare the even 
and odd contributions among themselves, they seem to form two independent geometric sequences as functions 
of $\l$. In other words, for sufficient large values of  $\ell$, we have (when grouping together either 
even or odd values of $\l$) $\ln E^{\rm rgd}_\ell \propto -\,\ell$,  $\ln J^{\rm rgd}_\ell \propto -\,\ell$.  
 This behavior is illustrated in Fig.~\ref{fig:q12} for the $q=1$ and $q=2$ cases. Note that such a simple decay law 
 $\propto \exp{(- c \, \l)}$ is suggested by the leading-order post-Newtonian (PN) expression of the various multipoles, namely
 $ h_\lm \propto n^{(\epsilon)}_\lm \, c_{\l + \epsilon}(\nu) \, \v^{\l+\epsilon} \exp(- \ii m \Phi)$ ,  where  $\epsilon = 0,1$ denotes the parity, $ n^{(\epsilon)}_\lm $ 
 is a parity-dependent numerical factor,  $c_{\l + \epsilon}(\nu)$ a mass-ratio dependent one,  $\v = r_\omega \Omega$ 
 a suitably defined (and resummed) azimuthal velocity, and $\Phi$ the orbital phase; see~\cite{Damour:2008gu}. 
 Previous EOB studies have shown that such ``Newtonian-order"  expressions (after a suitable resummation of the PN corrections) 
 stay numerically close to the full NR waveform essentially up to merger.  As the value of $\v$ at merger is still smaller than one (and
 as the orbital motion stays quasi-circular till merger), we can indeed expect a simple parity-dependent $\l$-exponential 
 decay of the multipolar contributions to the fluxes  $\dot{E}_\l ,  \dot{J}_\l$. [Actually, the $\l$ behavior of the fluxes results
 from combining various $\l$-dependent factors; in addition, one must take into account the effect of time-integration.
 Anyway, our phenomenological findings are useful independently of any precise analytical justification.]

 This leads us to introduce separate, even or odd, decay ratios, say (using the fact that, in the case at hand, $\ell_{\rm max}$ is even)
 \begin{align}
 r^E_e &= E^{\rm rgd}_{\ell_{\rm max}}/ E^{\rm rgd}_{\ell_{\rm max}-2}, \\
 r^E_o &=  E^{\rm rgd}_{\ell_{\rm max}-1}/E^{\rm rgd}_{\ell_{\rm max}-3},\\
 r^J_e &= J^{\rm rgd}_{\ell_{\rm max}}/ J^{\rm rgd}_{\ell_{\rm max}-2},\\
  r^J_o &=  J^{\rm rgd}_{\ell_{\rm max}-1}/J^{\rm rgd}_{\ell_{\rm max}-3}.
 \end{align}
 Here, we used as best estimates of the $E$ and $J$ decay ratios the largest-$\l$'s we can consider.  As we shall
 see later, the numerical values of the $E$ and $J$ decay ratios are found to be approximately equal: 
 $  r^E_e \approx  r^J_e \approx  r^E_o \approx  r^J_o $, but we did not use this approximate fact in our analysis. 

In view of this simple geometric-series decay of the multipolar contributions to the ringdown losses, we can improve
their $\lmax$-truncated estimates ~\eqref{eq:DeltaE}, ~\eqref{eq:DeltaJ}  by summing the two separate geometric series
representing the higher-$\l$ contributions. This leads to the following ``corrected'' estimate of the NR ringdown losses
(still in the case $q\leq 2$)
\begin{align}
 &E^{\rm rgd}_{\rm corr.} = \sum_{\ell =2}^{\infty}  E^{\rm rgd}_\l \nonumber\\
 &= E^{\rm rgd}_{(\ell_{\rm max})} +  E^{\rm rgd}_{\ell_{\rm max}}\,\frac{r^E_e}{1-r^E_e}\,+\, E^{\rm rgd}_{\ell_{\rm max}-1}\,\frac{r^E_o}{1-r^E_o}\,, \\
 &J^{\rm rgd}_{\rm corr.} = \sum_{\ell =2}^{\infty}  J^{\rm rgd}_\l \nonumber\\
 &=\,J^{\rm rgd}_{(\ell_{\rm max})} +  J^{\rm rgd}_{\ell_{\rm max}}\,\frac{r^J_e}{1-r^J_e}\,+\, J^{\rm rgd}_{\ell_{\rm max}-1}\,\frac{r^J_o}{1-r^J_o}\,.
\end{align}
The values of the ratios $r^{E,J}_{e,o}(q)$ we obtained are: (i)  for $q=1$; $r^E_e = 0.02908, r^E_o=0.03272$, $r^J_e = 0.02857, r^J_o=0.03286$ and (ii) for $q=2$; $r^E_e = 0.05467, r^E_o=0.05280$ , $r^J_e = 0.05249, r^J_o=0.05495$ . 

\begin{figure}[h]
 \includegraphics[width=0.45\textwidth]{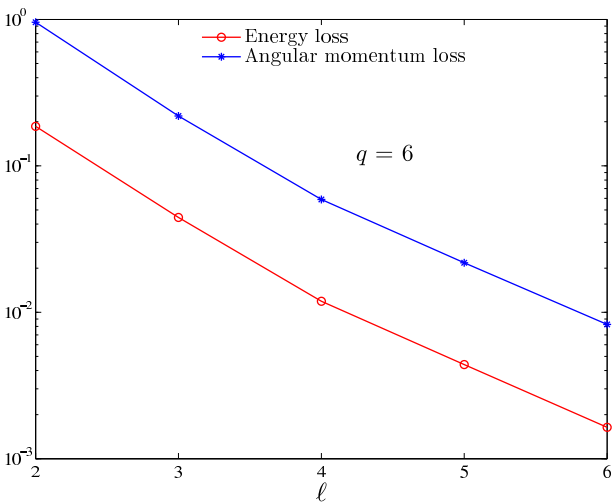}
    \caption{\label{fig:q6} Logarithms of  the $\l$-th multipolar contributions, $ E^{\rm rgd}_\ell$, $J^{\rm rgd}_\ell$,
    to the ringdown losses, as functions of $\ell$ for $q=6$,  exhibiting a uniform behavior for even and odd values of $\l$.} 
\end{figure}

Fig.~\ref{fig:q12}  showed that the pronounced even/odd difference present when $q=1$ 
is already much weaker when $q=2$. For larger values of $q$, namely $q\geq 3$, we found 
a uniform $\l$-behavior in that the even and odd partial contributions arrange themselves (for large enough $\l$) along 
a {\it common} geometric-series.  See, as an illustration, Fig.~\ref{fig:q6} for the $q=6$ case. 
In other words, there is no need to introduce separate even and odd ratios, and one can simply 
introduce a common ratio between $\l -1 $ and $\l $, say estimated from the highest possible value of $\l$,
\bea
 r^E = E^{\rm rgd}_{\ell_{\rm max}}/ E^{\rm rgd}_{\ell_{\rm max}-1} \, ,   \\
 r^J = J^{\rm rgd}_{\ell_{\rm max}}/ J^{\rm rgd}_{\ell_{\rm max}-1} \,    \, .
\eea
This common ratio would be related to the separate ones introduced above via $r_o=r_e=r^2$.
The corresponding improved estimate of the total ringdown losses is then obtained as
\bea
E^{\rm rgd}_{\rm corr.} = \ \sum_{\ell =2}^{\infty} E^{\rm rgd}_\l\,=\,E^{\rm rgd}_{(\ell_{\rm max})}  +  E^{\rm rgd}_{\ell_{\rm max}}\,\frac{r^E}{1-r^E}\,\,,  \\
J^{\rm rgd}_{\rm corr.} = \ \sum_{\ell =2}^{\infty} J^{\rm rgd}_\l\,=\,J^{\rm rgd}_{(\ell_{\rm max})}  +  J^{\rm rgd}_{\ell_{\rm max}}\,\frac{r^J}{1-r^J}\,\,.
\eea

The values of the ratios $r^{E,J}(q)$ we obtained for $q\geq 3$ are: (i)  for $q=3$;  $r^E =0.2918$, $r^J=0.2871$, (ii) for $q=4$; $r^E =0.3339$, $r^J=0.3315$, and (iii) for $q=6$; $r^E =0.3730$, $r^J=0.3795$. 

The final results we obtained by this method are displayed in  Table~\ref{tab:losses}. Let us recall that these results are
based: for $q=1$ on the results of~\cite{Scheel:2008rj} ,  and for $q=2,3,4,6$ on the results of~\cite{Buchman:2012dw}.
For the case $q=1$ we had data  up to $\ell_{\rm max}=8$ while for the other NR data we had data only
up to $\ell_{\rm max}=6$. In addition, we report in this Table the corresponding results  for the test-mass case $q=\infty$ 
(actually $q=10^3$), based on the results of ~\cite{Bernuzzi:2011aj}. For that case, we had data up to $\ell_{\rm max}=8$. 
In the table we indicate both the ``uncorrected'' values of the losses (up to $\ell_{\rm max}$ or  $\ell_{\rm max} -2$), 
and the corrected ones.

\begin{table*}[t]
  \caption{\label{tab:losses} NR losses of energy and angular momentum during ringdown, i.e. after the maximum of $h^{\rm NR}_{22}$.} 
  \begin{center}
    \begin{ruledtabular}
      \begin{tabular}{ccccc|cc}
        $\nu$      &   $q$      &   $\ell_\text{max}$ &  $ E^{\rm rgd}/\nu^2$   &   $ J^{\rm rgd}/\nu^2$ & $E^{\rm rgd}_{\rm corr.}/\nu^2$ & $J^{\rm rgd}_{\rm corr.}/\nu^2$\\
        \hline 
$0$                       &  $\infty$  &  $8$ &  $0.267836$  &  $1.51821$  &  $0.270011$  & $1.53037$ \\
$0$                       &  $\infty$  &  $6$ &  $0.262383$  &  $1.48763$  &                   &    \\
$0.1224...$     &  $6$       &  $6$ &  $0.248517$  &  $1.26535$  &  $0.249492$  & $1.27041$ \\
$0.16$                 &  $4$       &  $6$ &  $0.254961$  &  $1.24448$  &  $0.255392$  & $1.24651$ \\
$0.1875$             &  $3$       &  $6$ &  $0.263139$  &  $1.24301$  &  $0.263344$  & $1.24393$ \\
$0.\bar{2}$     &  $2$       &  $6$ &  $0.276747$  &  $1.24870$  &  $0.276807$  & $1.24898$ \\
$0.25$                &  $1$       &  $6$ &  $0.293199$  &  $1.27188$  &                   &    \\
$0.25$                &  $1$       &  $8$ &  $0.293204$  &  $1.27190$  &  $0.293204$  & $1.27190$
 \end{tabular}
  \end{ruledtabular}
\end{center}
\end{table*}

The ringdown losses displayed  in  Table~\ref{tab:losses} have been scaled by a factor $\nu^2$ because each 
multipolar waveform $\Psi_{\lm}(t)= (R/M) h_{\lm}/\sqrt{(\l+2)(\l+1) \l (\l-1)}$ contains a factor $\nu$ which is conveniently
factored out. In our previous work ~\cite{Damour:2007cb}, we had proposed to approximate the $\nu$ dependence of
the ringdown losses by assuming that this $\nu^2$ scaling holds for any value of $\nu$. This would correspond to
assuming that the ratios $E^{\rm rgd}_{\rm corr.}/\nu^2, J^{\rm rgd}_{\rm corr.}/\nu^2$ do not depend on $\nu$.
As we see on  Table~\ref{tab:losses}, this assumption is approximately correct. For instance, as $\nu$ varies
between $0$ (test-mass case) and $1/4$ (equal-mass case),  $E^{\rm rgd}_{\rm corr.}/\nu^2$ varies  between
$0.27$ and $0.29$, passing through a minimum $\simeq 0.25$. Similarly, $J^{\rm rgd}_{\rm corr.}/\nu^2$ varies  between
$1.53$ and $1.27$, passing through a minimum $\simeq 1.24$. Compared to the $\nu^2$ scaling, these are rather
mild variations. Let us also note that the effect of higher multipoles is fractionally more important as $\nu$ becomes small.
This explains why our current best estimates of the corrected values of  $E^{\rm rgd}_{\rm corr.}/\nu^2, J^{\rm rgd}_{\rm corr.}/\nu^2$ for $\nu \to 0$, namely $0.27001$ and $1.53037$ significantly differ from the corresponding values,
$0.2448$ and $1.3890$ we had used in~\cite{Damour:2007cb}, which were based on summing only up to
$\lmax=4$ multipoles.

Before discussing in more detail the relative importance of the various multipolar orders to the
results  in  Table~\ref{tab:losses}, let us mention that the final, corrected values listed there (for a
sample of $q$ values) can be approximately represented by the following polynomials in $\nu$:
\begin{align}\label{eq:fitlosses}
E^{\rm rgd}_{\rm corr.}/\nu^2 &= 2.0337 \,\nu^2\,-0.41783\,\nu\,+\,0.27005 \, , \nonumber \\
J^{\rm rgd}_{\rm corr.}/\nu^2 &= 8.3205\,\nu^2\,-3.1076\,\nu\,+\,1.5298\, .
\end{align}
These $\nu$-dependent fits reproduce the values of the {\it unscaled} ringdown losses 
$E^{\rm rgd}_{\rm corr.}, J^{\rm rgd}_{\rm corr.}$
corresponding to the above-listed $q$-sample of  scaled losses modulo  differences smaller than
$- 4.1 \times 10^{-5}$ for  $E^{\rm rgd}_{\rm corr.}$ (fit minus exact difference reached when $q=2$), and 
$+ 1.5 \times 10^{-4}$ for  $J^{\rm rgd}_{\rm corr.}$ (fit minus exact difference reached when $q=3$).
In the following, we shall use these fits only when we will need an estimate of the ringdown losses
in the range $0< \nu <0.1224$ where we do not have sufficient NR data to estimate them by the
method explained above.

As it can happen that one does not have at hand NR data covering higher multipolar contributions, 
it is useful to study  which fraction of the total ringdown losses is due to the dominant $\lm = 2 2$ (even) 
quadrupolar  waveform, and which fraction is due to the subdominant contributions (i.e., in the present 
context, the subdominant quadrupolar waveforms $\lm = 2 1$ and $\lm = 2 0$, as well as the higher 
multipolar orders $\l \geq 3$). Let us define the fractional contribution of subdominant multipoles 
to the ringdown losses as
\begin{align}
f_E\,&=\,\frac{E^{\rm rgd}_{\rm corr.} - E^{\rm rgd}_{22}}{E^{\rm rgd}_{\rm corr.}}\, ,\\
f_J\,&=\,\frac{J^{\rm rgd}_{\rm corr.} - J^{\rm rgd}_{22}}{J^{\rm rgd}_{\rm corr.}}\, ,
\end{align}
so that, e.g., $E^{\rm rgd}_{\rm corr.}=  E^{\rm rgd}_{22}/( 1 - f_E)$, etc.

\begin{figure}[h]
 \includegraphics[width=0.45\textwidth]{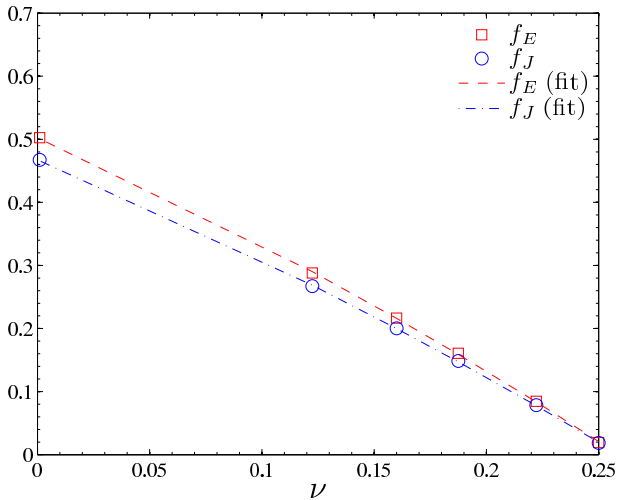}
    \caption{\label{rat22} Fraction of the total ringdown losses of energy and angular momentum (resp. $f_E$ and $f_J$) 
    due to higher multipoles ($\lm \neq 2 2$) as functions of $\nu$ for NR data and the test-mass values. 
    Both corrected values and the fits~Eqs.~\eqref{fE} and \eqref{fJ} are plotted.} 
\end{figure}

We give in  Table~\ref{tab:deltaE22}  the values of the fractions $f_E$ and $f_J$ for the $q$ values we studied here.
The $\nu$ dependence of  $f_E$ and $f_J$ is illustrated in  Fig.~\ref{rat22}. This figure plots both the points listed in
Table~\ref{tab:deltaE22}, and the following polynomial fits to $f_E$ and $f_J$
\begin{align}
f_E &= -1.51473\,\nu^2\,-1.54739\,\nu\,+\,0.502328\,,\label{fE}\\
f_J &= -1.32120\,\nu^2\,-1.45965\,\nu\,+\,0.467261\,.\label{fJ}
\end{align}

\begin{table}[h]
  \caption{\label{tab:deltaE22} 
  Relative contribution of all subdominant terms (i.e. different from $\ell=m=2$) to the total ringdown 
  losses of energy and angular momentum: $f_Q=1-Q^{\rm rgd}_{22}/ Q^{\rm rgd}_{\rm corr.}$ with $Q$ = $E$ or $J$.} 
  \begin{center}
    \begin{ruledtabular}
      \begin{tabular}{cccc}
        $\nu$      &   $q$      &   $f_E$   &   $f_J$ \\
        \hline 
        \hline
$0$              &  $\infty$   &  0.502552   &  0.467411\\
$0.1224...$      &  $6$        &  0.287962   &  0.267059\\
$0.16$           &  $4$        &  0.216489   &  0.200202\\
$0.1875$         &  $3$        &  0.160672   &  0.148506\\
$0.\bar{2}$      &  $2$        &  0.0846702   &  0.0785165\\
$0.25$           &  $1$        &  0.0193487  &  0.0185868
 \end{tabular}
  \end{ruledtabular}
\end{center}
\end{table}

Table~\ref{tab:deltaE22}  and  Fig.~\ref{rat22} show that the fractions  $f_E$  and   $f_J$ {\it strongly depend} on $\nu$.
More precisely, while the  $\ell=m=2$ waveform
contributes alone most of the ringdown losses when one is close to the equal-mass case $\nu \simeq 0.25$, it contributes
only about half of the total losses when one is close to the extreme-mass ratio case $\nu \ll 1$.  However, we hope that
our fits (\ref{fE}),  (\ref{fJ}) can provide a useful alternative to using many NR multipoles, by simply dividing the
$\lm =2 2$ NR ringdown contributions by the corresponding factors $1- f_E(\nu)$  or   $1- f_J(\nu)$.

\section{Dynamical state at merger: NR/EOB  comparison}

In this Section we shall compare the analytical  EOB prediction for the dynamical state of the system at merger
to the corresponding  result that can be inferred from NR data.  Here, like in any EOB work, the word ``analytical"
is to be understood in the sense that the EOB formalism provides a set of analytical equations of motion for
the dynamical state of the system. These equations are, however, too complex to be solved exactly, and
one resorts to standard (Runge-Kutta-type) numerical tools to obtain their solutions. Due to the sensitivity to the EOB initial conditions, in most cases the EOB results given below with six significant digits are actually
accurate within fractional errors of order $10^{-5}$.

From the NR side, if we start from the
{\it final} values of the energy and angular momentum of the system (i.e. the mass $M_f$ and angular momentum
$J_f$ of the final BH resulting from the coalescence of the two BH's), we can estimate the energy and angular momentum
of the system at the moment of merger, say $E^{\rm mrg}_{\rm NR}$,  and $J^{\rm mrg}_{\rm NR}$, by adding to
 $M_f$ and $J_f$ the corresponding ringdown losses:
 \bea
 E^{\rm mrg}_{\rm NR}:= M_f ^{\rm NR} +  E^{\rm rgd}_{\rm NR} \, , \\
  J^{\rm mrg}_{\rm NR}:= J_f ^{\rm NR} +  J^{\rm rgd}_{\rm NR} \, .
 \eea
Here, for clarity, we added a subscript NR to the (corrected) NR ringdown losses estimated in the previous
Section. We added also a superscript NR to  $M_f$ and $J_f$. Indeed, we took for them the values directly
indicated in the original NR papers~\cite{Scheel:2008rj,Buchman:2012dw}, without using the various
approximate fitting formulas for them that have been 
proposed in the literature~\cite{Damour:2007cb,Rezzolla:2007rz,Tichy:2008du,Barausse:2009uz,Pan:2011gk,Barausse:2012qz}. 

\begin{table*}[t]
\caption{\label{tab:mg} Total energy and angular momentum at the time of merger 
for NR and EOB (see the text).}
\begin{center}
\begin{ruledtabular}
\begin{tabular}{cc|cc|cc}
$\nu$ & $q$ & $E^{\rm mrg}_{\rm NR}$ & $E^{\rm mrg}_{\rm EOB}$ &  $J^{\rm mrg}_{\rm NR}$ & $J^{\rm mrg}_{\rm EOB}$\\
\hline
\hline
$0.1224\dots$  & $6$   & $0.98921(5)$ & $0.989495$ & $0.38075(10)$ & $0.380534$\\
$0.16$         & $4$   & $0.98446(2)$ & $0.984978$ & $0.48292(10)$ & $0.482790$\\
$0.1875$       & $3$   & $0.98054(1)$ & $0.981218$ & $0.55371(2)$ & $0.553486$\\
$0.\bar{2}$    & $2$   & $0.97491(2)$ & $0.975784$ & $0.63773(4)$ & $0.637243$\\
$0.25$         & $1$   & $0.96995(2)$ & $0.971156$ & $0.70114(4)$ & $0.702198$
\end{tabular}
\end{ruledtabular}
\end{center}
\end{table*}

 From the EOB side, the EOB code computes at any moment (on the EOB time axis) the total energy
 $E_{\rm EOB}(t) = H(t)$ and total angular momentum   $J_{\rm EOB}(t) = \nu p_{\varphi}(t)$ of the binary system.
 To compute the values of  $E_{\rm EOB}(t)$ and $J_{\rm EOB}(t)$ at (EOB) merger we just need to know (from
 the EOB code itself) what moment $\tmg_{\rm EOB}$ on the EOB time axis corresponds to ``merger''.
 The (conventional) definition we gave above of ``merger''  is equally meaningful within the NR context and within the
 EOB one.  In other words, the EOB merger moment is simply the (EOB) time $\tmg_{\rm EOB}$  when the 
 modulus of the $\lm = 2 2$ EOB strain waveform $h_{22}^{\rm EOB}(t)$ reaches its maximum.
 The (EOB) merger energy and angular momentum are then: 
 \bea
 E^{\rm mrg}_{\rm EOB}  := E_{\rm EOB}(\tmg_{\rm EOB})\, , \\
  J^{\rm mrg}_{\rm EOB}  :=  J_{\rm EOB}(\tmg_{\rm EOB}) \, .
 \eea
 Here, we have been using the fact that, within the EOB formalism, there is  well-defined link between the
 dynamical time, and the (retarded) time used to express the waveform. The EOB waveform
 used here to find $\tmg_{\rm EOB}$ is the  inspiral-plus-plunge  waveform defined in Eq.~(16) of~\cite{Damour:2012ky},
 including the  next-to-quasi-circular correction factor ${\hat h}^{\rm NQC}_\lm$ of  Eq.~(27) there. As the maximum
 of $|h_{22}|$ occurs {\it before} the (EOB) moment  $\teobLR$ where the orbital frequency $\Omega$ peaks 
 (and where one attaches the sum of quasi-normal modes modelling the ringdown), the determination of EOB merger
 does not depend on this quasi-normal-mode attachment procedure. On the other hand, the determination of EOB merger
 is not a purely analytical procedure that directly follows from analytically known first principles. It is a procedure 
 that depends both on the analytically known part of EOB, and on the NR-informed improvements leading to the current 
 EOB[NR] formalism. Notably, the determination of the precise value of  $\tmg_{\rm EOB}$ involves the knowledge of
 the parameters $a_i$  entering the modulus of the  next-to-quasi-circular correction factor ${\hat h}^{\rm NQC}_{2 2}$ to the $\lm =22$ waveform.
 These parameters have been determined in~\cite{Damour:2012ky} (as functions of $\nu$) by extracting some non-perturbative 
 information from NR waveforms (namely the shape of the NR modulus $|h^{\rm NR}_{22}(t^{\rm NR}) |$ near a certain NR instant 
 $\tnrLR$, located just after $\tmg_{\rm NR}$).
 
 The comparison between the pure NR quantities   $E^{\rm mrg}_{\rm NR}$,  $J^{\rm mrg}_{\rm NR}$ and the
 corresponding EOB[NR] ones  $E^{\rm mrg}_{\rm EOB}$,  $J^{\rm mrg}_{\rm EOB}$ is presented in Table~\ref{tab:mg}.
 The NR/EOB agreement is quite good, both for $E^{\rm mrg}$ and for $J^{\rm mrg}$. It is essentially at the  $10^{-3}$ (fractional)  
 level, which is remarkable when one thinks that the physically complicated merging two-black-hole system we 
 are considering is represented (at the time $\tmg_{\rm EOB} < \teobLR$), in the EOB formalism,  by means of 
 a rather simple analytic Hamiltonian (corresponding to a two point-mass system). 
 To put this level of NR/EOB agreement in perspective let us note that the purely NR check of the
 conservation of energy and angular momentum (between the initial state and the final state of a $q=1$ system) 
 done in~\cite{Reisswig:2009rx}
 yielded fractional agreements of order  $| E^{\rm in}_{\rm NR} - E^{\rm rad}_{\rm NR} -E^{\rm f}_{\rm NR}|/E^{\rm f}_{\rm NR} = 7.6 \times 10^{-5}$  and  $| J^{\rm in}_{\rm NR} - J^{\rm rad}_{\rm NR} -J^{\rm f}_{\rm NR}|/J^{\rm f}_{\rm NR} = 5.5 \times 10^{-4}$.

  Let us also note that the EOB merger energies are systematically  {\it larger} than the NR ones. 
 This (small) systematic difference might simply be due to the fact (already emphasized in~\cite{Damour:2011fu}) that,
 in using the consequences of simple conservation laws for $E$ and $J$, we have neglected the ``Schott contributions"
 to $E$ and $J$. This issue has been recently studied within the EOB formalism~\cite{Bini:2012ji}. The latter reference
 showed that (with the choice of radiation reaction adopted in the present EOB formalism~\cite{Damour:2012ky}),
  only the energy had a Schott contribution $E^{\rm Schott}_{\rm EOB}$, and that this contribution was small during the
 inspiral and was {\it negative} (because proportional to the radial momentum).
  This result actually means that, when defining the NR merger energy as above,
 it should be compared to the sum  $E^{\rm mrg}_{\rm EOB} +  E^{\rm Schott}_{\rm EOB} <  E^{\rm mrg}_{\rm EOB}$,
 rather than to $E^{\rm mrg}_{\rm EOB}$ alone.  This correction (if it remains small until merger) might therefore 
 further improve the agreement between EOB and NR. However,  the current analytical result derived in~\cite{Bini:2012ji}
 for  $E^{\rm Schott}_{\rm EOB}$ is only given in a non-resummed PN form, which cannot be
 meaningfully applied to the moment of merger. Indeed, when trying to use it to determine its effect on the above
 NR/EOB comparison, we realized that the PN correction factor was, numerically, of the form
 $ 1- a/c^2 + b/c^4$ with both $a/c^2$ and $b/c^4$ positive and of order one at merger (and with $a/c^2 \approx 1+ b/c^4$, suggesting
 that the whole correction factor is significantly smaller than one).  We leave to future work the study of possible
 {\it resummations} of the Schott-energy results of~\cite{Bini:2012ji}.  It is only when such (well-tested)
 resummations are constructed that it will become meaningful to take into account deviations from the
 naive conservation-laws used above. 
 
\section{Final black hole state}

Let us now compare the NR results for the final dynamical state of the system, $M_f^{\rm NR}$, $J_f^{\rm NR}$, 
to the results obtained by subtracting from the  EOB predictions for the dynamical state of the system at merger 
the subsequent (NR-computed) ringdown losses. In other words, we define, on the EOB side
\bea
\label{eobef}
 M_f^{\rm EOB[NR]} := E^{\rm mrg}_{\rm EOB} -  E^{\rm rgd}_{\rm NR} \, , \\
J_f^{\rm EOB[NR]} := J^{\rm mrg}_{\rm EOB} -  J^{\rm rgd}_{\rm NR} \, .
\label{eobjf}
\eea
 This comparison is essentially equivalent to the one done in the previous Section, being just the
other side of the same coin. However, it is interesting in itself to formulate the NR/EOB comparison in terms of final
state quantities (rather than merger ones) both because the final-state parameters are of more physical interest
than the merger ones, and because this comparison updates the one made in our previous work~\cite{Damour:2007cb}.

Table~\ref{tab:kp} summarizes the NR/EOB comparison of final (mass-)energy $M_f$ and angular momentum $J_f$,
as well as the corresponding dimensionless Kerr parameter $\a_f:=J_f/M_f^2$.  The fractional level of EOB/NR 
agreement is again (as a consequence of the merger agreement above) at the $10^{-3}$ level, as shown in 
Table~\ref{tab:frd}. For instance, in the equal-mass case (where the larger number of
available multipoles, $\lmax=8$, allowed a better determination of the ringdown losses) the absolute agreement between 
$\a_{\rm NR} = 0.68646(4)$ (where the notation $(4)$ indicates the NR probable error on the last digit) and  $\a_{\rm EOB[NR]} = 0.68590$ is   $\a_{\rm EOB[NR]} -  \a_{\rm NR}= - 5.6 \times 10^{-4}$.

\begin{table*}[t]
\caption{\label{tab:kp} Properties of the final BH: comparison between NR and EOB[NR] predictions.}
\begin{center}
\begin{ruledtabular}
\begin{tabular}{cc|cc|cc|cc}
 $\nu$ &  $q$ & $M_f^{\rm NR}$ &  $M_f^{\rm EOB[NR]}$ & $J_f^{\rm NR}$ & $J_f^{\rm EOB[NR]}$ & $\a_{f}^{\rm NR}$ &
$\a_{f}^{\rm EOB[NR]}$\\
\hline
\hline
$0.1224\dots$ & $6$  & $0.98547(5)$  & $0.985754$ & $0.36171(10)$ & $0.361486$  & $0.37245(10)$ &  $0.372009$\\
$0.16$        & $4$  & $0.97792(2)$  & $0.978440$ & $0.45100(10)$ & $0.450880$  & $0.47160(10)$ &  $0.470969$ \\
$0.1875$      & $3$  & $0.97128(1)$  & $0.971960$ & $0.50997(2)$  & $0.509754$  & $0.54058(2)$  &  $0.539590$ \\
$0.\bar{2}$   & $2$  & $0.96124(2)$  & $0.962115$ & $0.57605(4)$  & $0.575565$  & $0.62344(4)$  &  $0.621785$ \\
$0.25$        & $1$  & $0.95162(2)$  & $0.952830$ & $0.62164(4)$  & $0.622704$  & $0.68646(4)$  &  $0.685884$
\end{tabular}
  \end{ruledtabular}
\end{center}
\end{table*}

\begin{table}[h]
  \caption{\label{tab:frd} Fractional differences between EOB[NR] and NR for the mass, the angular momentum and the dimensionless Kerr parameter of the final BH, defined as $\Delta Q_f/Q_f=(Q_f^\text{EOB[NR]}-Q_f^\text{NR})/Q_f^\text{NR}$ with $Q=M,J$ or $\a$.} 
  \begin{center}
    \begin{ruledtabular}
      \begin{tabular}{cc|ccc}
        $\nu$      &   $q$  &    $\Delta M_f/M_f$  &  $\Delta J_f/J_f$ & $\Delta \a_f/\a_f$  \\
        \hline
        \hline
$0.1224\dots$  &  $6$   & $+2.88\,\times\,10^{-4}$  &  $-6.06\,\times\,10^{-4}$  &  $-1.18\,\times\,10^{-3}$\\
$0.16$        &  $4$   & $+5.32\,\times\,10^{-4}$  &  $-2.77\,\times\,10^{-4}$  &  $-1.34\,\times\,10^{-3}$\\
$0.1875$      &  $3$   & $+7.00\,\times\,10^{-4}$  &  $-4.35\,\times\,10^{-4}$  &  $-1.83\,\times\,10^{-3}$\\
$0.\bar{2}$ &  $2$   & $+9.10\,\times\,10^{-4}$  &  $-8.37\,\times\,10^{-4}$  &  $-2.65\,\times\,10^{-3}$\\
$0.25$        &  $1$   & $+1.27\,\times\,10^{-3}$  &  $+1.73\,\times\,10^{-3}$  &  $-8.20\,\times\,10^{-4}$
\end{tabular}
  \end{ruledtabular}
\end{center}
\end{table}

 Up to here we have only considered the NR/EOB comparison for the mass ratios $q=(1,\,2,\, 3,\, 4,\, 6)$ 
 simulated with the SpEC code~\cite{Scheel:2008rj,Buchman:2012dw}.  Such mass ratios are of prime 
 importance for the upcoming ground-based gravitational-wave detectors. However,  larger mass ratios 
 can be relevant within more general contexts, such as astrophysical studies of merging BH's, or  future space-based 
 gravitational-wave detectors. One of the advantages of having an accurate analytical understanding of the
 dynamics of coalescing BHs, such as the one provided by EOB theory, is that one can predict quantities of
 direct physical interest (such as $M_f$ and $\a_f$) for mass ratios that are difficult to simulate with good accuracy.
 In addition EOB theory provides predictions for the continuous range of $\nu$ values, thereby
 accomplishing both an interpolation and an extrapolation of the information gathered by NR simulations of 
 a discrete sample of $q$ values. Indeed, by inserting now in Eqs.~\eqref{eobef},\eqref{eobjf} the polynomial 
 fits~\eqref{eq:fitlosses} of the ringdown losses (instead of the NR-computed losses), we can transform
 calculations of the dynamical state at merger into estimates of the dynamical quantities
 $M_f$, $J_f$ and $\a_f$, of the final BH. Before reporting  the results of such EOB[NR] calculations for generic values
 of $\nu$, let us compare what they give in the three specific higher-$q$ cases that have been recently numerically
 simulated by  Lousto and collaborators~\cite{Lousto:2010qx,Lousto:2010ut,Lousto:2010xk}.  The results of this 
 comparison are displayed in Table~\ref{tab:lousto}. There is a good agreement between the two sets of results.
 For instance, the absolute (rather than fractional) differences between  the values of  $\a_f$ are all of the order of $10^{-4}$.
 It is difficult to put these differences in perspective because  Refs.~\cite{Lousto:2010qx,Lousto:2010ut,Lousto:2010xk}
 do not provide estimates of the probable uncertainty on their (numerically very challenging) results.

 In Table \ref{tab:eobnr} we give more results of the EOB[NR] predictions for the final masses and spin-parameters of coalescing
 BH's, as a function of the symmetric mass ratio $\nu$.  Though, by downloading the freely available EOBIHES code~\cite{eobihes}
 [and using Eqs.~\eqref{eq:fitlosses},\eqref{eobef},\eqref{eobjf}] any reader can compute such values, it might be useful
 to encapsulate our results by means of some fitting formulas.  Let us first give accurate ``local fits'' for the $\nu$ dependence of  
 $M_f$ and $\a_f$ interpolating between the Caltech-Cornell-CITA NR results~\cite{Scheel:2008rj,Buchman:2012dw}.  
 We found that the following polynomials
 \begin{align}
& \a_f^{\rm NR fit}(\nu)  = \sqrt{12} \nu \nonumber\\
                     & + \left(41.59 \nu^3 - 37.3036 \nu^2 + 13.9195 \nu - 4.6713\right)\nu^2, \\
& M_f^{\rm NR fit}(\nu)  = 1 + \left(\sqrt{\dfrac{8}{9}}-1\right) \nu \nonumber\\
&                      +(-2.962 \nu^2 + 0.765 \nu -0.5514) \nu^2\,,
 \end{align}
 reproduce , within their quoted probable errors,  the NR results, for the sample of $\nu$ values corresponding
  to the NR $q$ sample $q=(1,\,2,\,3,\,4,\,6)$. More precisely, the differences  
  $| \a_f^{\rm NR fit}(\nu) - \a_f^{\rm NR }(\nu)|$ and    $| M_f^{\rm NR fit}(\nu) - M_f^{\rm NR }(\nu)|$
  are smaller than $1.8 \times 10^{-6}$ and   $6.9 \times 10^{-6}$ respectively.
  These fits are more accurate than the ones given in~\cite{Damour:2007cb,Rezzolla:2007rz,Tichy:2008du,Pan:2011gk},
  and can therefore be advantageously used in the EOB code to estimate the characteristics of the final BH,
  when one is interested in $q$ values within the range $1 \leq q \lesssim 6$. [Let us correct in passing a small misprint
  in Eq.~(32) of ~\cite{Damour:2012ky} : its left-hand-side should read $a_f/M_f$.]

\begin{table}[t]
  \caption{\label{tab:lousto} Final mass and dimensionless final Kerr parameter: comparison with 
  estimates from Lousto {\it et al.}~\cite{Lousto:2010qx,Lousto:2010ut,Lousto:2010xk}.} 
  \begin{center}
    \begin{ruledtabular}
      \begin{tabular}{cc|cc|cc}
        $\nu$      &   $q$  &    $M_f^{\rm NR}$  &  $M_f^{\rm EOB[NR]}$   & $\a_f^{\rm NR}$  &  $\a_f^{\rm EOB[NR]}$  \\
        \hline 
        \hline
$0.0098\dots$     &  $100$   &  0.999382    &   0.999379   &  0.0333  &  0.0334146 \\
$0.05859\dots$   &   $15$   &  0.99493     &   0.994897   &  0.18875 &  0.188834 \\
$0.08264\dots$   &   $10$   &  0.99174     &   0.991899   &  0.2603  &  0.260176        
\end{tabular}
  \end{ruledtabular}
\end{center}
\end{table}

\begin{table}[t]
  \caption{\label{tab:eobnr} EOB[NR] predictions for the mass and dimensionless Kerr parameter of the final BH calculated using  
EOB results at merger together with the fits~\eqref{eq:fitlosses} of the losses.} 
  \begin{center}
    \begin{ruledtabular}
      \begin{tabular}{cc|cc}
        $\nu$      &   $q$  &  $M_f^{\rm EOB[NR]}$  &  $\a_f^{\rm EOB[NR]}$ \\
        \hline 
        \hline
$0.02$    &  $47.979...$   & $0.998627$   &  $0.0672678$   \\
$0.04$    &  $22.956...$   & $0.996862$   &  $0.131453$   \\
$0.06$    &  $14.598...$   & $0.994736$   &  $0.193092$  \\
$0.08$    &  $10.403...$   & $0.992255$   &  $0.252480$  \\
$0.1$     &  $7.8729...$   & $0.989408$   &  $0.309825$
\end{tabular}
  \end{ruledtabular}
\end{center}
\end{table}
On the other hand, if one is interested in having simple analytical representations of the variation of   $M_f$ and $\a_f$
over the full range $0< \nu \leq 0.25$, it might be useful to have fits that approximately encapsulate our new 
EOB[NR] results (a sample of which is displayed in Tables~\ref{tab:kp}, \ref{tab:lousto}, \ref{tab:eobnr}, 
as well as \ref{tab:compoldeob} below).  For this case, we found that the 
following ``global fits'' for  the $\nu$ dependence of   $M_f$ and $\a_f$ might be adequate:
\begin{align}
& \a_f^{\rm EOB[NR] fit}(\nu)  = \sqrt{12} \nu \nonumber\\
                     & + \left(255.5 \nu^3 - 146.5 \nu^2 + 31.467 \nu - 5.5534\right)\nu^2, \\
& M_f^{\rm EOB[NR] fit}(\nu)  = 1 + \left(\sqrt{\dfrac{8}{9}}-1\right) \nu \nonumber\\
&                      +(-4.815 \nu^2 + 1.315 \nu -0.579) \nu^2\,,
\end{align}
Note, however, that the latter fits are less accurate (compared to current NR data) than the former ones, when
evaluated at the  NR $q$ sample $q=1,2,3,4,6$.  More precisely,  for these values of $q$ the maximum differences
 $| \a_f^{\rm NR fit}(\nu) - \a_f^{\rm NR }(\nu)|$ and    $| M_f^{\rm NR fit}(\nu) - M_f^{\rm NR }(\nu)|$
  are only smaller than $1.4 \times 10^{-3}$ and   $3.7 \times 10^{-4}$ respectively.

\begin{figure}[t]
 \includegraphics[width=0.45\textwidth]{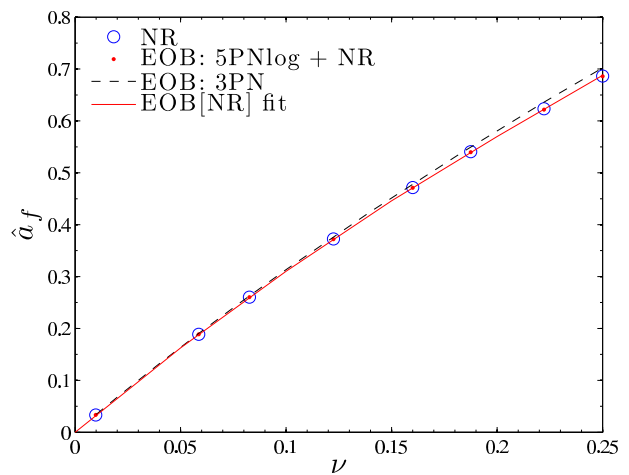}
    \caption{\label{fig:compoldeob} Dimensionless Kerr parameter as a function of $\nu$: comparison 
    of NR data~\cite{Scheel:2008rj,Buchman:2012dw}, current (5PNlog+NR) EOB[NR] predictions and 
    their global fit, and earlier (3PN accurate) EOB estimates (joined plot) \cite{Damour:2007cb}.}
\end{figure}

Let us finally discuss the extent to which our new EOB[NR] results for   $M_f$ and $\a_f$ represent an
improvement with respect to our previous work~\cite{Damour:2007cb}, which used a purely analytical 
(3PN- accurate) EOB formalism, together with  $\nu^2$-scaled ringdown losses, inferred from an early 
test-mass calculation~\cite{Nagar:2006xv}.  A quantitative comparison of the results of~\cite{Damour:2007cb} 
with the present ones is provided in Table~\ref{tab:compoldeob}.  The results of~\cite{Damour:2007cb} 
displayed here correspond to the then a priori preferred Damour-Gopakumar-type~\cite{Damour:2006tr}  radiation reaction. As,  indeed, this (non-Keplerian) type of
resummation of the leading-order term in the waveform (and  radiation reaction) plays a crucial role in the
resummed EOB radiation reaction  \cite{Damour:2008gu} incorporated in the current EOB formalism, it is appropriate
to contrast them here with the new EOB results. Note, in particular, that,
in the $q=1$ case,  the old (3PN-accurate) EOB result of~\cite{Damour:2007cb} 
was  $\a^{\rm EOB[3PN]}_f = 0.7023$. The  EOB/NR agreement  has been improved by a factor 
$(0.7023-0.68646)/(0.68590-0.68646) \simeq -30$. This improvement in the EOB/NR agreement 
is further illustrated in Fig.~\ref{fig:compoldeob} which compares the 3PN-accurate earlier predictions of~\cite{Damour:2007cb} both to the predictions of the current (5PNlog+NR)  EOB[NR] predictions, and to the recent NR results . This confirms
the trend discussed in~\cite{Damour:2007cb} (which notably compared 2PN-accurate versions of EOB to 3PN-accurate ones): 
the addition of more accurate physics in the EOB formalism leads, at the end, to a better agreement with NR results. 

\begin{table}[t]
  \caption{ \label{tab:compoldeob}From left to right the columns report: the symmetric mass ratio $\nu$, 
   the final dimensionless angular momentum $\a_f$ and the final mass $M_f$ for both the most recent
  (5PN with logs, NR-completed) EOB formalism of Ref.~\cite{Damour:2012ky} and the earlier 3PN-accurate (without NR completion)  EOB formalism
  of Ref.~\cite{Damour:2007cb}. }
  \begin{ruledtabular}
    \begin{tabular}{c|cc|cc}
      $\nu$  & $M_f^{\rm 5PNlog+NR}$ & $M_f^{\rm 3PN}$ & $\a_f^{\rm 5PNlog+NR}$ & $\a_f^{\rm 3PN}$ \cr
      \hline
      \hline
      0.01   &   $0.999366$     &  0.9994 & $0.00340769$  & 0.0341 \\
      0.02   &   $0.998627$     &  0.9986 & $0.0672678$     & 0.0675 \\
      0.04   &   $0.996862$     &  0.9969 & $0.131453$     & 0.1322 \\
      0.06   &   $0.994736$     &  0.9949 & $0.193092$     & 0.1946 \\
      0.08   &   $0.992255$     &  0.9925 & $0.252480$     & 0.2549  \\
      0.10   &   $0.989408$     &  0.9898 & $0.309825$      & 0.3134  \\
      0.14   &   $0.982538$     &  0.9834 & $0.418981$      & 0.4251 \\
      0.1825 &   $0.973223$     &  0.9751 & $0.527487$      & 0.5370  \\      
      0.2227 &   $0.961923$     &  0.9659 & $0.622867$      & 0.6373 \\
      0.25   &   $0.952830$     &  0.9589 & $0.685884$      & 0.7023
    \end{tabular}
  \end{ruledtabular}
\end{table}

\section{Conclusions}

The main results of the present investigation are:

\begin{enumerate}

\item[(i)]  We presented a new quantitative test of the ability of  EOB  theory to accurately describe the dynamics
of coalescing binary BH's. This test goes beyond recent such tests \cite{LeTiec:2011bk,Damour:2011fu} in that
it deeply probes the strong-gravitational-field regime. We found that EOB theory predicts values for the energy
and angular momentum of the binary system at the moment of merger which agree, to the per mil level, with
corresponding values that were inferred from recent accurate NR computations  \cite{Scheel:2008rj,Buchman:2012dw}.
In a sense, this aspect of our work is an extension of the recent study of the 
gauge-invariant energy-angular-momentum relation
\cite{Damour:2011fu} from the inspiral regime down to the fully relativistic regime of merger.

  \item[(ii)] Our work is also an update of a previous EOB study of the final mass and spin of a coalescing binary 
  BH  \cite{Damour:2007cb}.  While the latter study used a purely analytical (3PN-accurate) EOB formalism,
  completed by $\nu^2$-scaled test-mass estimates of the ringdown losses, we have used here a recently developed
  (analytically more complete, NR-informed) EOB formalism \cite{Damour:2012ky,eobihes}, together with refined,
  NR-based estimates of the ringdown losses (based on Refs. \cite{Scheel:2008rj,Buchman:2012dw}).  We then find
  that this leads to a much increased agreement between NR and EOB (e.g. by a factor $\simeq 30$ for the final
  spin parameter). This  confirms the trend found in \cite{Damour:2007cb}:
   the addition of more accurate physics in the EOB formalism leads,
 at the end, to a better agreement with NR results.  In our opinion, this shows that  EOB theory, beyond giving
a phenomenologically useful description of the gravitational-wave emission of coalescing binaries, does also provide
a simple, but physically correct description of the dynamical state of a binary BH system, down to the moment of merger.
  
  \item[(iii)]  Another potentially useful outcome of our study  concerns  the exponential decay with multipolar
  order $\l$ of the integrated losses of energy and angular momentum during ringdown. We showed how
  this decay can be used to improve the sum over $\l$. Essentially, we proposed a Richardson extrapolation
  of the sum to infinite $\lmax$. We also gave explicit $\nu$-dependent fits of the ratio between the dominant
  $\lm=22$ contribution and the sum of all the subdominant ones. Such fits might be useful for estimating the
  total losses in cases where NR simulations recorded only the dominant   $\lm=22$ multipole.

  \item[(iv)] From the practical point of view, our work also provides several simple fitting formulas that might
  be useful in various contexts.  For instance, we provide both global and local fits for the mass-ratio dependence
  of the final mass and spin parameters of (non spinning) coalescing binaries, which incorporate the most recent,
  accurate NR results (together, in the case of the global fit, with the results of our present EOB[NR] study).
  
\end{enumerate}

\begin{acknowledgments}
We thank the Caltech-Cornell-CITA collaboration for making available the numerical waveform  data 
of  Refs.~\cite{Scheel:2008rj,Buchman:2012dw}; we are grateful to Luisa Buchman and Harald Pfeiffer
for informative communications about their simulations. We also thank Sebastiano Bernuzzi for his  continuous
help in improving the EOBIHES code~\cite{eobihes}  that we used in this work. LV thanks IHES for hospitality
during the devopment of this work.
\end{acknowledgments}



\begin{thebibliography}{99}
\bibitem{Buonanno:1998gg} A.~Buonanno and T.~Damour,
Phys.\ Rev.\ D {\bf 59}, 084006 (1999)
\bibitem{Buonanno:2000ef} A.~Buonanno and T.~Damour,
Phys.\ Rev.\ D {\bf 62}, 064015 (2000)
\bibitem{Damour:2000we} T.~Damour, P.~Jaranowski and G.~Schaefer,
Phys.\ Rev.\ D {\bf 62}, 084011 (2000)
\bibitem{Damour:2001tu} T.~Damour,
Phys.\ Rev.\ D {\bf 64}, 124013 (2001)
\bibitem{Buonanno:2005xu} 
  A.~Buonanno, Y.~Chen and T.~Damour,
  Phys.\ Rev.\ D {\bf 74}, 104005 (2006)
  [gr-qc/0508067].
\bibitem{Damour:2008gu} 
  T.~Damour, B.~R.~Iyer and A.~Nagar,
  Phys.\ Rev.\ D {\bf 79}, 064004 (2009)
  [arXiv:0811.2069 [gr-qc]].

\bibitem{Damour:2012mv} 
  T.~Damour,
  arXiv:1212.3169 [gr-qc].

\bibitem{Pretorius:2005gq} 
  F.~Pretorius,
  Phys.\ Rev.\ Lett.\  {\bf 95}, 121101 (2005)
  [gr-qc/0507014].

\bibitem{Campanelli:2005dd} 
  M.~Campanelli, C.~O.~Lousto, P.~Marronetti and Y.~Zlochower,
  Phys.\ Rev.\ Lett.\  {\bf 96}, 111101 (2006)
  [gr-qc/0511048].

\bibitem{Baker:2005vv} 
  J.~G.~Baker, J.~Centrella, D.~-I.~Choi, M.~Koppitz and J.~van Meter,
  Phys.\ Rev.\ Lett.\  {\bf 96}, 111102 (2006)
  [gr-qc/0511103].


\bibitem{Pretorius:2007nq} 
  F.~Pretorius,
  arXiv:0710.1338 [gr-qc].
  
\bibitem{Pfeiffer:2012pc} 
  H.~P.~Pfeiffer,
  Class.\ Quant.\ Grav.\  {\bf 29}, 124004 (2012)
  [arXiv:1203.5166 [gr-qc]].
  
\bibitem{Mroue:2013xna} 
  A.~H.~Mroue, M.~A.~Scheel, B.~Szilagyi, H.~P.~Pfeiffer, M.~Boyle, D.~A.~Hemberger, L.~E.~Kidder and G.~Lovelace {\it et al.},
  arXiv:1304.6077 [gr-qc].

  
\bibitem{Buonanno:2006ui} 
  A.~Buonanno, G.~B.~Cook and F.~Pretorius,
  Phys.\ Rev.\ D {\bf 75}, 124018 (2007)
  [gr-qc/0610122].
  
\bibitem{Pan:2007nw} 
  Y.~Pan, A.~Buonanno, J.~G.~Baker, J.~Centrella, B.~J.~Kelly, S.~T.~McWilliams, F.~Pretorius and J.~R.~van Meter,
  Phys.\ Rev.\ D {\bf 77}, 024014 (2008)
  [arXiv:0704.1964 [gr-qc]].

\bibitem{Buonanno:2007pf} 
  A.~Buonanno, Y.~Pan, J.~G.~Baker, J.~Centrella, B.~J.~Kelly, S.~T.~McWilliams and J.~R.~van Meter,
  Phys.\ Rev.\ D {\bf 76}, 104049 (2007)
  [arXiv:0706.3732 [gr-qc]].
  
\bibitem{Damour:2007vq} 
  T.~Damour, A.~Nagar, E.~N.~Dorband, D.~Pollney and L.~Rezzolla,
  Phys.\ Rev.\ D {\bf 77}, 084017 (2008)
  [arXiv:0712.3003 [gr-qc]].

\bibitem{Damour:2008te} 
  T.~Damour, A.~Nagar, M.~Hannam, S.~Husa and B.~Bruegmann,
  Phys.\ Rev.\ D {\bf 78}, 044039 (2008)
  [arXiv:0803.3162 [gr-qc]].
  
\bibitem{Damour:2009kr} 
  T.~Damour and A.~Nagar,
  Phys.\ Rev.\ D {\bf 79}, 081503 (2009)
  [arXiv:0902.0136 [gr-qc]].

\bibitem{Buonanno:2009qa} 
  A.~Buonanno, Y.~Pan, H.~P.~Pfeiffer, M.~A.~Scheel, L.~T.~Buchman and L.~E.~Kidder,
  Phys.\ Rev.\ D {\bf 79}, 124028 (2009)
  [arXiv:0902.0790 [gr-qc]].
  
\bibitem{Pan:2009wj} 
  Y.~Pan, A.~Buonanno, L.~T.~Buchman, T.~Chu, L.~E.~Kidder, H.~P.~Pfeiffer and M.~A.~Scheel,
  Phys.\ Rev.\ D {\bf 81}, 084041 (2010)
  [arXiv:0912.3466 [gr-qc]].


\bibitem{Pan:2011gk} 
  Y.~Pan, A.~Buonanno, M.~Boyle, L.~T.~Buchman, L.~E.~Kidder, H.~P.~Pfeiffer and M.~A.~Scheel,
  Phys.\ Rev.\ D {\bf 84}, 124052 (2011)
  [arXiv:1106.1021 [gr-qc]].




\bibitem{Taracchini:2012ig} 
  A.~Taracchini, Y.~Pan, A.~Buonanno, E.~Barausse, M.~Boyle, T.~Chu, G.~Lovelace and H.~P.~Pfeiffer {\it et al.},
  Phys.\ Rev.\ D {\bf 86}, 024011 (2012)
  [arXiv:1202.0790 [gr-qc]].
  
  \bibitem{Damour:2012ky} 
  T.~Damour, A.~Nagar and S.~Bernuzzi,
Phys. Rev. D {87}, 084035 (2013)

\bibitem{Damour:2006tr} 
  T.~Damour and A.~Gopakumar,
  Phys.\ Rev.\ D {\bf 73}, 124006 (2006)
  [gr-qc/0602117].


\bibitem{Damour:2007cb} 
  T.~Damour and A.~Nagar,
  Phys.\ Rev.\ D {\bf 76}, 044003 (2007)
  [arXiv:0704.3550 [gr-qc]].
  
\bibitem{LeTiec:2011bk} 
  A.~Le Tiec, A.~H.~Mroue, L.~Barack, A.~Buonanno, H.~P.~Pfeiffer, N.~Sago and A.~Taracchini,
  Phys.\ Rev.\ Lett.\  {\bf 107}, 141101 (2011)
  [arXiv:1106.3278 [gr-qc]].


\bibitem{Damour:2011fu} 
  T.~Damour, A.~Nagar, D.~Pollney and C.~Reisswig,
  Phys.\ Rev.\ Lett.\  {\bf 108}, 131101 (2012)
  [arXiv:1110.2938 [gr-qc]].
  
  \bibitem{eobihes}
  The corresponding ``EOBIHES" code developed in Ref.~\cite{Damour:2012ky}
  and written in {\tt MatLab} is freely available at http://eob.ihes.fr.
  
\bibitem{Scheel:2008rj} 
  M.~A.~Scheel, M.~Boyle, T.~Chu, L.~E.~Kidder, K.~D.~Matthews and H.~P.~Pfeiffer,
  Phys.\ Rev.\ D {\bf 79}, 024003 (2009)
  [arXiv:0810.1767 [gr-qc]].

  
\bibitem{Buchman:2012dw} 
  L.~T.~Buchman, H.~P.~Pfeiffer, M.~A.~Scheel and B.~Szilagyi,
  Phys.\ Rev.\ D {\bf 86}, 084033 (2012)
  [arXiv:1206.3015 [gr-qc]].
  
\bibitem{Lousto:2010xk} 
  C.~O.~Lousto and Y.~Zlochower,
  Phys.\ Rev.\ D {\bf 83}, 024003 (2011)
  [arXiv:1011.0593 [gr-qc]].

  
\bibitem{Lousto:2010qx} 
  C.~O.~Lousto, H.~Nakano, Y.~Zlochower and M.~Campanelli,
  Phys.\ Rev.\ D {\bf 82}, 104057 (2010)
  [arXiv:1008.4360 [gr-qc]].

\bibitem{Lousto:2010ut} 
  C.~O.~Lousto and Y.~Zlochower,
  Phys.\ Rev.\ Lett.\  {\bf 106}, 041101 (2011)
  [arXiv:1009.0292 [gr-qc]].
  
\bibitem{Bernuzzi:2011aj} 
  S.~Bernuzzi, A.~Nagar and A.~Zenginoglu,
  Phys.\ Rev.\ D {\bf 84}, 084026 (2011)
  [arXiv:1107.5402 [gr-qc]].
  
\bibitem{Rezzolla:2007rz} 
  L.~Rezzolla, E.~Barausse, E.~N.~Dorband, D.~Pollney, C.~Reisswig, J.~Seiler and S.~Husa,
Phys.\ Rev.\ D {\bf 78}, 044002 (2008)
[arXiv:0712.3541 [gr-qc]].

\bibitem{Tichy:2008du} 
  W.~Tichy and P.~Marronetti,
Phys.\ Rev.\ D {\bf 78}, 081501 (2008)
[arXiv:0807.2985 [gr-qc]].
  
\bibitem{Barausse:2009uz} 
  E.~Barausse and L.~Rezzolla,
  Astrophys.\ J.\  {\bf 704}, L40 (2009)
  [arXiv:0904.2577 [gr-qc]].

\bibitem{Barausse:2012qz} 
  E.~Barausse, V.~Morozova and L.~Rezzolla,
  Astrophys.\ J.\  {\bf 758}, 63 (2012)
  [arXiv:1206.3803 [gr-qc]].
   

  
\bibitem{Reisswig:2009rx} 
  C.~Reisswig, N.~T.~Bishop, D.~Pollney and B.~Szilagyi,
  Class.\ Quant.\ Grav.\  {\bf 27}, 075014 (2010)
  [arXiv:0912.1285 [gr-qc]].


\bibitem{Bini:2012ji} 
  D.~Bini and T.~Damour,
  Phys.\ Rev.\ D {\bf 86}, 124012 (2012)
  [arXiv:1210.2834 [gr-qc]].
  
\bibitem{Nagar:2006xv} 
  A.~Nagar, T.~Damour and A.~Tartaglia,
  Class.\ Quant.\ Grav.\  {\bf 24}, S109 (2007)
  [gr-qc/0612096].

\end{thebibliography}
\end{document}